\documentstyle[preprint,prl,aps]{revtex}
\begin{document}
\draft

\title{Quantum Corrections for Generalized Partition 
Functions}
\author{L. R. Evangelista, L. C. Malacarne}
\address{Departamento de F\'\i sica - Universidade 
Estadual de Maring\'a \\
Av. Colombo 5790, 87020-900 Maring\'a, PR, Brazil}
\author{R. S. Mendes\footnote{Permanent address:
{Departamento de F\'\i sica - Universidade Estadual 
de Maring\'a \\
Av. Colombo 5790, 87020-900 Maring\'a, PR, Brazil}}}
\address{Centro Brasileiro de Pesquisas F\'\i sicas, \\
Rua Xavier Sigaud 150, 22290-180 Rio de Janeiro, RJ
 Brazil}
\date{\today}
\maketitle

\begin{abstract}
The classical limit  for  generalized partition functions 
is obtained using coherent states. In this framework it
is presented a general
procedure to obtain all the corrections to the classical limit.
In particular, the first and second order quantum corrections 
are worked out explicitly, and the classical 
limit for the Tsallis thermostatistics is determined. 
The results of this work generalize the ones obtained by 
E. Wigner (Phys. Rev. 40 (1932) 749) for usual statistical
 mechanics.
\end{abstract}

\pacs{PACS number(s): 05.20.-y, 05.70.-a, 05.30.Ch }

\date{\today}

\section{Introduction}

Long-range interactions are present in many physical contexts.
The nature of these interactions can be connected to spatial 
and temporal dependencies. For 
instance, anomalous diffusion\cite{1,1a,1b,1c}, astrophysics with 
long-range (gravitational) interactions\cite{2,2a,2b,2c,2d}, 
some magnetic systems\cite{3,3a,3b}, some surface tension
questions\cite{4,4a}.
In such cases we have a nonextensive behavior. 
This fact indicates that usual 
thermodynamics and statistical mechanics deserves
some changes. 
Thus, it is important to generalize 
the concepts based in the usual thermostatistics. 
In this direction, it has been recently analyzed the Legendre 
transform structure\cite{5,6} and stability conditions\cite{6} 
in a very general context. 
The present work is devoted for kind of study. 
More precisely, the purpose of this work is to  obtain the
 classical
limit  
and its quantum corrections for arbitrary partition functions. 

It is important to obtain the classical limit and its corrections 
because, in some cases,
it is easier to calculate the classical partition function
than the quantum one.
This problem was solved first for the usual statistical mechanics 
 by Wigner\cite{6a} using the 
Wigner functions, and by Kirkwood\cite{7} employing the Bloch
 equations
(see also reference\cite{7a,7b}).
The procedure employed by 
Kirkwood can not be applied easily for generalized 
partition functions.
On the other hand, the procedure used by Wigner can be 
in principle extended for arbitrary partition functions, 
but in this work we prefer to employ coherent states\cite{7c}. 
By using coherent states we classify and give the 
prescription to calculate all the possible 
corrections for the classical partition function. 
Employing this prescription we obtain 
explicitly the first two corrections.

The general results obtained in this work can be
in principle  applied to an arbitrary  
thermostatistics.
In particular,  they are used in the context of 
Tsallis thermostatiscs.
This thermostatistics was proposed in order to cover 
nonextensive systems (long-range microscopic memory, 
long-range forces, fractal space time).
Thus, the Tsallis 
thermostatistics\cite{8} has  interesting proprieties 
and has been applied in many 
situations, like for instance, self-gravitating systems
\cite{9,9a,10},
two-dimensional-like turbulence\cite{10}, 
L\'evy-like\cite{12,12a,12b,12c,12d} and 
correlated-like\cite{13,13a,13b,13c} anomalous diffusion, 
solar neutrino problem\cite{14}, linear response theory
\cite{15} and
magnetic systems\cite{16,16a,16b,16c}.

\section{Quantum corrections}

In this work we consider a partition function for 
a very general statistical mechanics,
\begin{eqnarray}
\label{1}
Z = Tr f(\hat{H}),
\end{eqnarray}
where $Tr$ represents the trace, $\hat H$ is the Hamiltonian 
of the 
system, and $f(x)$ is the generalization of the exponential
 function 
of the usual thermostatistics.

To calculate the classical limit of (\ref{1}) 
and its corrections it is convenient 
to employ coherent states. 
In particular, two basic properties of the 
coherent states are important for our calculations. 
The first one is the 
definition of coherent states, namely
\begin{eqnarray}\label{2}
\hat a_n \mid z > = z_n \mid z >\; ,
\end{eqnarray}
where
\begin{eqnarray}\label{3}
\hat a_n = \left(\sqrt{\frac{m_n \omega}{2\hbar}} \hat q_n +
\frac i{\sqrt{2 m_n \omega\hbar}} \hat p_n  \right ) \; ,
\end{eqnarray}
\begin{eqnarray}\label{4}
z_n = \left(\sqrt{\frac{m_n \omega}{2\hbar}}  q_n +
\frac i{\sqrt{2 m_n \omega\hbar}} p_n  \right ) \; ,
\end{eqnarray}
and $\mid z> = \prod_{n=1}^{N}\mid z_n >$. 
In this expression for $\mid z >$, $N$ is the dimension of
the configuration space, and each $\mid z_n > $ is 
normalized, $ < z_n \mid z_n > = 1$. 
Moreover,  $q_n$ and $p_n$ are respectively the classical
values for coordinate and momentum $n$.
The other property is the overcompleteness relation,
\begin{eqnarray}
1=\int \prod_{n=1}^{N} \frac{dp_n dq_n}{2\pi\hbar} 
\; {\mid z><z\mid}\; .
\label{5}
\end{eqnarray}
Furthermore, employing the previous notation, we use $\hat A$ to
 represent
an operator and $A$ to represent its classical analog.

To obtain the corrections to the classical partition function we
 basically
expand $f(\hat H)$ around its classical value, $f(H)$. 
More precisely, we write
$\hat H(\hat p_n, \hat q_n)$ as
 $\hat H(p_n +\delta \hat  p_n,q_n+ \delta\hat q_n)$ and express
 $f(\hat
H)$
as a power series in $\delta \hat H$,
\begin{eqnarray}\label{6}
f(H+\delta\hat H) =f(H)+ \frac{df}{dH}\; \delta\hat H +
 \frac12 \frac{d^2f}{dH^2} (\delta\hat H)^2 + ...  \; ,
\end{eqnarray}
where
\begin{eqnarray}\label{7}
\delta\hat H = H(p_n+\delta\hat p_n,q_n+\delta\hat q_n) 
- H(p_n,q_n) \; ,
\end{eqnarray}
\begin{eqnarray}\label{8}
\hat H = H(\hat p_n,\hat q_n)= \sum_{n=1}^{N} 
\frac{\hat p_n^2}{2m_n} + 
V(\hat q_n) \; ,
\end{eqnarray}
$\delta\hat q_n =\hat q_n -q_n$ and $\delta\hat p_n = 
\hat p_n- p_n$.
Note that in the present analysis we are considering  
a Hamiltonian that depends only on the spatial coordinates and 
its corresponding momenta.

By using the relation (\ref{5}) we can now express the partition 
function
as
\begin{eqnarray}
\label{9}
Z = \int \prod_{n=1}^{N} \frac{dp_n dq_n}{2\pi\hbar} 
\; {<z\mid f(H+\delta\hat H) \mid z >} \; ,
\end{eqnarray}
and to calculate it we employ the expansion presented above. 
The first
term of this expansion gives the classical partition function,
\begin{eqnarray}
\label{10}
Z_0 = \int \prod_{n=1}^{N} \frac{dp_n dq_n}{2\pi\hbar}\; 
 f(H) \; .
\end{eqnarray}

The corrections to (\ref{10}) can be classified using 
a simple but important result, i.e.
\begin{eqnarray}\label{11}
{<z\mid \delta\hat X_1 ... \delta\hat X_u \mid z>}
\propto \hbar^{{u}/{2}} \; ,
\end{eqnarray}
where $\delta\hat X_n$ can be $\delta\hat q_n$ or $\delta\hat p_n$.
In fact, the expression (\ref{11}) comes directly from the 
definition of $\delta\hat q_n$ and $\delta\hat p_n$,
\begin{eqnarray}
\label{12}
 \delta\hat q_n=\sqrt{\frac{\hbar}{2m_n \omega}}
\left( \delta\hat a_n^{\dagger} + \delta\hat a_n\right )\; 
\end{eqnarray}
and
\begin{eqnarray}
\label{13}
 \delta\hat p_n=i \; \sqrt{\frac{m_n \omega \hbar}{2}}
\left( \delta\hat a_n^{\dagger} - \delta\hat a_n\right )\; , 
\end{eqnarray}
where $\delta\hat a_n = \hat a_n - z_n$ and
$\delta\hat a_n^{\dagger} = \hat a_n^{\dagger} - z_n^*$.
Furthermore, when $u$ is an odd number a direct calculation
 leads to
$<z\mid \delta\hat X_1 ... \delta\hat X_u \mid z > =0 $.
 Therefore, the 
corrections to (\ref{10}) are proportional to integer powers of
$\hbar$.
Thus, the partition function can be written as
\begin{eqnarray}\label{13a}
Z=Z_0 + \hbar Z_1 + \hbar ^2 Z_2 + ...\; \; .
\end{eqnarray}
For instance, the first and second quantum corrections,
 $\hbar Z_1$ and
$\hbar Z_2$, comes respectively from the second and fourth
 order expansions
in $\delta\hat q_k$ and $\delta \hat p_k$.

Let us now proceed to calculate the first quantum correction.
 First we
note,
by means of equation (\ref{2}), (\ref{12}) and (\ref{13}), that
\begin{eqnarray}
\label{14}
<z\mid \left\{ 
\matrix {
\delta\hat q_n \delta \hat q_k \cr
\delta\hat q_n \delta \hat p_k \cr  
\delta\hat p_n \delta \hat q_k \cr 
\delta\hat p_n \delta \hat p_k \cr }
\right \} \mid z > = 
\hbar \left \{ \matrix {
 1/({m_n \omega}) \cr 
i \cr
-i \cr 
m_n \omega \cr }
\right \} \frac{\delta_{nk}}{2}\; .
\end{eqnarray}
Moreover, from (\ref{6}), (\ref{7}) and (\ref{8}) these mean 
values are present only in
$\delta\hat H$ and $(\delta\hat H)^2$. Consequently, by using 
the equations
(\ref{7}),(\ref{8}) and (\ref{14}) we conclude that
\begin{eqnarray}\label{15}
{<z\mid \delta\hat H\mid z>}=
\frac{\hbar \omega}{2} \sum_{n=1}^{N} \left[  \frac 12  
+ \frac{1}{2m_n \omega^2}\frac{\partial ^2 V}{\partial q_n^2} 
 \right]
+{\cal O}(\hbar ^2)
\end{eqnarray}
and
\begin{eqnarray}\label{16}
{<z\mid (\delta\hat H)^2\mid z>}=
\frac{\hbar \omega}{2} \sum_{n=1}^{N}
 \left[  \frac {p_n^2}{m_n} 
+ \frac{1}{m_n \omega^2}\left(\frac{\partial  V}
{\partial q_n}\right )^2 
\right]
+{\cal O}(\hbar ^2) \; . 
\end{eqnarray}
It follows directly from (\ref{6}), (\ref{9}), (\ref{13a}),
 (\ref{15}) and (\ref{16})
that
\begin{eqnarray}\label{17}
Z_1 = \frac \omega 4 \int \prod _{n=1}^{N} \frac{dq_n dp_n}
{2\pi\hbar}
\sum_{j=1}^{N} \left\{ \left ( \frac{df}{dH} + 
\frac {p_j^2}{m_j }  \frac{d^2 f}{dH^2} \right ) +
\frac{1}{m_j \omega^2} 
\left [ \frac{\partial ^2 V}{\partial q_j^2} \frac{df}{dH} + 
\left(\frac{\partial  V}{\partial q_j}\right )^2
\frac{d^2 f}{dH^2} \right ]  \right \} \; .
\end{eqnarray}
This expression can be written in a more compact form. 
In fact, after some derivations we verify that 
\begin{eqnarray}\label{18}
Z_1 = \frac \omega 4 \int \prod _{n=1}^{N} \frac{dq_n dp_n}
{2\pi\hbar}
\sum_{j=1}^{N} \left[
\frac{\partial}{\partial p_j} \left( p_j \frac{df}{dH} \right) 
+ \frac{1}{m_j \omega^2}\frac{\partial}{\partial q_j}
\left(\frac{\partial V}{\partial q_j} \frac{df}{dH}
\right)
\right] \; .
\end{eqnarray}
Furthermore, the expression (\ref{18}) is more convenient 
than (\ref{17}) for future discussions.

The calculation of the second quantum correction, $\hbar^2 Z_2$, 
is similar to the previous one. 
In fact, to obtain the analogous formula to (\ref{14}), 
\begin{eqnarray}
\label{18a}
<\delta X_n \delta X_k \delta X_s \delta X_u > &=&
- m_{n}^{-\sigma_{n}/2} m_{k}^{-\sigma_{k}/2} 
m_{s}^{-\sigma_{s}/2} m_{u}^{-\sigma_{u}/2} 
(i\omega)^{-(\sigma_{n} + \sigma_{k} + \sigma_{s} + \sigma_{u})/2} 
\nonumber \\
&\times& 
\sigma_n [ \sigma_s \delta_{nk} + 
\sigma_k ( \delta_{ku} \delta_{sn} +\delta_{nu} \delta_{ks})] \; ,
\end{eqnarray}
it is necessary only the equations 
(\ref{2}), (\ref{12}) and (\ref{13}).
In the expression (\ref{18a}) it was used a compact notation
suggested in the Eq.~(\ref{11})
because there are sixteen combinations of $\delta q_j$ and
$\delta p_j$. 
In this notation
$\delta X_j =  \hbar^{1/2} i^{(1-\sigma_j)/2} 
(m_j \omega)^{-\sigma_j /2} 
(\delta a_{j}^{\dag} + \sigma_j \delta a) $ is equal to
$\delta q_j$ when $\sigma_j =1$ and equal to
$\delta p_j$ when $\sigma_j =-1$. 
By using the expressions (\ref{14}) and (\ref{18a}) 
we can obtain  after some calculation
the contributions proportional to $\hbar$ and $\hbar^2 $ in 
$<z\mid (\hat H)^n \mid z>$, i. e.
\begin{eqnarray}
\label{18b}
<z\mid \delta \hat H \mid z> =
\frac{\hbar \omega}{2} \sum_{n=1}^{N} \left[  \frac 12  
+ \frac{1}{2m_n \omega^2}\frac{\partial ^2 V}{\partial q_n^2}  
\right]
+\frac{\hbar^2}{ 4 \omega^2 } 
\sum_{nk} \frac{1}{8m_n m_k} 
\frac{\partial^4 V}{\partial q^2_n \partial_k^2}
+{\cal O}(\hbar ^3) \; , 
\end{eqnarray}
\begin{eqnarray}
\label{18bb}
< \! &z&\mid (\delta \hat H)^2 \mid z> =
\frac{\hbar \omega}{2} \sum_{n=1}^{N} \left[  \frac {p_n^2}{m_n} 
+ \frac{1}{m_n \omega^2}\left(\frac{\partial  V}
{\partial q_n}\right )^2 
\right] + 
\frac{\hbar^2}{4} \sum_{j,k=1}^{N} 
\left \{ \frac 1{\omega^2 m_j m_k}
\left[ \frac{\partial V}{\partial q_j} 
\frac{\partial^3 V}{\partial q_j\partial q_k^2}\right. \right. 
\nonumber \\ 
&+& \left. \left. \frac 14 \frac{\partial^2 V}{\partial q_j^2} 
\frac{\partial^2 V}{\partial q_k^2} 
+\frac 12 \left(\frac{\partial^2 V}{\partial q_j
\partial q_k}\right)^2
\right]
+ \left( -\frac{\delta_{jk}}{m_j} + \frac 1{2m_j} \right)  
\frac{\partial ^2 V}{\partial q_k^2}  
+ 
\omega^2 
\left( \frac 14 + \frac{\delta_{jk}}{2} \right) \right\}
+{\cal O}(\hbar ^3) \; ,
\end{eqnarray}
\begin{eqnarray}
\label{18bbb}
<\! &z& \mid (\delta \hat H)^3 \mid z> =
\frac{3\hbar^2}{4} \sum_{j,k=1}^{N}
 \left \{ \frac 1{\omega^2 m_j m_k}
\left[ \frac 12
\left( \frac{\partial V}{\partial q_j}\right)^2
\frac{\partial^2 V}{\partial q_k^2}
+ \frac{\partial V}{\partial q_j}
\frac{\partial V}{\partial q_k}
\frac{\partial^2 V}{\partial q_j\partial q_k}
\right]
+  \frac{p_j^2}{2m_j m_k} 
 \frac{\partial ^2 V}{\partial q_k^2} \right.  \nonumber \\ 
&-& \left. \frac{p_j p_k}{3 m_j m_k}
 \frac{\partial ^2 V}{\partial q_j \partial q_k} 
+ \frac 1{2m_j} \left( \frac{\partial  V}{\partial q_j}\right )^2 
\left( \delta_{kk} -\frac 23 \right) 
+  \omega^2 \frac{p_j^2}{2m} (\delta_{kk} +2)
\right\} + {\cal O}(\hbar ^3) \; ,
\end{eqnarray}
and
\begin{eqnarray}
\label{18bbbb}
<z\mid (\delta \hat H)^4 \mid z> &=&
{3\hbar^2} \sum_{j,k=1}^{N} \left \{ \frac 1{4\omega^2 m_j m_k}
\left( \frac{\partial V}{\partial q_j}\right)^2
\left(\frac{\partial V}{\partial q_k}\right)^2
+ \frac{p_j^2}{2m_j m_k}\left( \frac{\partial  V}
{\partial q_k}\right )^2 
+\omega^2 \frac{p_j^2}{2m_j}\frac{p_k^2}{2m_k}
\right\}\nonumber \\
 &+&  {\cal O}(\hbar ^3)
\; .
\end{eqnarray}

It is convenient now to group  the terms of the
 previous expressions 
with the same power in $\omega$. 
Thus, the expression for $Z_2$ can be written as 
\begin{eqnarray}\label{19}
Z_2 = \int \prod _{n=1}^{N} \frac{dq_n dp_n}{2\pi\hbar}
\left(
\frac{1}{w^2}I_1 + I_2 + w^2 I_3
\right) \; , 
\end{eqnarray}
where
\begin{eqnarray}\label{20}
I_1&=&\frac 18 \sum_{j,k=1}^{N} \frac 1{m_j m_k} \left\{
\left(\frac 14 \frac{\partial^4 V}{\partial q_j^2\partial q_k^2}
\right)\frac{df}{dH}
+ \left[ \frac{\partial V}{\partial q_j} 
\frac{\partial^3 V}{\partial q_j\partial q_k^2} 
+ \frac 14 \frac{\partial^2 V}{\partial q_j^2} 
\frac{\partial^2 V}{\partial q_k^2}
+
 \frac 12 \left(\frac{\partial^2 V}{\partial q_j\partial q_k}
\right)^2
\right] \frac{d^2f}{dH^2} \right. \nonumber \\
&+& \left. \left[ \frac 12
\left( \frac{\partial V}{\partial q_j}\right)^2
\frac{\partial^2 V}{\partial q_k^2}
+ \frac{\partial V}{\partial q_j}
\frac{\partial V}{\partial q_k}
\frac{\partial^2 V}{\partial q_j\partial q_k}
\right]
\frac{d^3f}{dH^3}
+ \left[ \frac 14
\left( \frac{\partial V}{\partial q_j}\right)^2
\left(\frac{\partial V}{\partial q_k}\right)^2
\right]\frac{d^4 f}{dH^4}
\right\}\; ,
\end{eqnarray}
 
\begin{eqnarray}\label{21}
I_2&=& \frac 18 \sum_{j,k=1}^{N}  \left\{ 
\left[ \left( -\frac{\delta_{jk}}{m_j} + \frac 1{2m_j} \right)  
\frac{\partial ^2 V}{\partial q_k^2} \right]
\frac{d^2f}{dH^2}
+ \left[ \frac{p_j^2}{2m_j m_k} 
 \frac{\partial ^2 V}{\partial q_k^2} 
-\frac{p_j p_k}{3 m_j m_k}
 \frac{\partial ^2 V}{\partial q_j \partial q_k} 
\right. \right. \nonumber \\
&+& \left. \left.
 \frac 1{2m_j} \left( \frac{\partial  V}{\partial q_j}\right )^2 
\left( \delta_{kk} -\frac 23 \right) 
\right]
\frac{d^3f}{dH^3}
+ \left[ 
\frac{p_j^2}{2m_j m_k}\left( \frac{\partial  V}
{\partial q_k}\right )^2 
\right]\frac{d^4 f}{dH^4}
\right\}\; , 
\end{eqnarray}
and
\begin{eqnarray}\label{22}
I_3=\frac 18 \sum_{j,k=1}^{N} \left\{ 
\left( \frac 14 + \frac{\delta_{jk}}{2} \right)
\frac{d^2f}{dH^2}
+ \left[ \frac{1}{2m} p_j^2 (\delta_{kk} +2)
\right]
\frac{d^3f}{dH^3}
+ \left( 
\frac{p_j^2}{2m_j}\frac{p_k^2}{2m_k}\right)\frac{d^4 f}
{dH^4}\right\} \; .
\end{eqnarray}
As in the $Z_1$ case it is more convenient, 
for the following discussions, to write 
$I_1$, $I_2$ and $I_3$ in a more compact form,
\begin{eqnarray}\label{23}
I_1&=&\frac 1{32} \sum_{j,k=1}^{N}
\frac{\partial ^3}{\partial q_j\partial q_k^2}
\left( \frac 1{m_j m_k} \frac{\partial V}{\partial q_j}
\frac{df}{dH}\right)\; ,
\end{eqnarray}
\begin{eqnarray}\label{24}
I_2&=&-\frac14 \sum_{k=1}^{N} \frac{1}{m_k}
\left( \frac{\partial^2 V}{\partial q_k^2}\right)
\frac{d^2f}{dH^2}
- \frac 1{6} \sum_{j,k=1}^{N}\left[ 
\frac{p_j}{m_j} \frac{ p_k}{m_k} 
\frac{\partial^2 V}{\partial q_j\partial q_k}
+ \frac{1}{m_j m_k} \left( \frac{\partial V}{\partial q_k}
\right)^2\right]
\frac{d^3 f}{dH^3}
\nonumber \\
&-&\frac 18 \sum_{k=1}^{N} 
\left(\frac{ p_k}{m_k} \frac{\partial V}{\partial q_k} 
\right)^2
\frac{d^4f}{dH^4} 
+ \frac18 \sum_{j,k=1}^{N} 
\frac{\partial}{\partial q_k}\frac{\partial}{\partial p_j}
\left[ \left(\frac {p_k}{ m_k} \frac{\partial V}{\partial q_j}+
\frac { p_j}{2 m_j} \frac{\partial V}{\partial q_k}
\right)\frac{d^2f}{dH^2}
\right]
\; ,
\end{eqnarray}
and
\begin{eqnarray}\label{25}
I_3= \frac 1{32} \sum_{j=1}^{N}  \frac{\partial}{\partial p_j}
\left[ p_j \frac{d^2f}{dH^2}+ \sum_{k=1}^{N} \frac{\partial}
{\partial p_k}
\left(p_j p_k \frac{d^2f}{dH^2}\right) \right] \; . 
\end{eqnarray}
This general procedure can be extended to other orders
 in $\hbar$ to 
obtains further quantum corrections, but we prefer to
 discuss the previous
results. 

In $Z_1$ and $Z_2$ are present terms which depends of the 
frequency $\omega$ employed in the definition of the 
coherent states. Note that these terms are surface terms.
However, a classical limit can not 
depend on the particular choice of the coherent states 
basis. 
This apparent contradiction can be eliminated if
we consider that in a macroscopic body the surface effects 
can be neglected when we compare them with the bulk contributions.
By using this consideration we conclude that the 
$\hbar Z_1$ contribution to the 
partition function can be neglected, and only $I_2$ is
relevant in the computation of $\hbar^{2} Z_2$. 
Therefore, when  these considerations are employed we verify 
directly that 
\begin{eqnarray}
\label{z}
Z &=& \int \prod_{n=1}^{N} \frac{dp_n dq_n}{2\pi\hbar} 
\; f(H)+ \hbar^2 \int \prod_{n=1}^{N} \frac{dp_n dq_n}{2\pi\hbar}
\left \{ -\frac14 \sum_{k=1}^{N}\frac 1{m_k} 
\left( \frac{\partial^2 V}{\partial q_k^2}\right)
\frac{d^2f}{dH^2} \right. \nonumber \\ 
&-& \left. \frac 1{6} \sum_{j,k=1}^{N}\left[ 
\frac{p_j}{m_j} \frac{ p_k}{m_k} 
 \frac{\partial^2 V}{\partial q_j\partial q_k}
+ \frac 1{m_j m_k}\left( \frac{\partial V}{\partial q_k}\right)
^2\right]
\frac{d^3 f}{dH^3}
-\frac 18 \sum_{k=1}^{N} 
\left( \frac{p_k}{m_k} \frac{\partial V}{\partial q_k} \right)^2
\frac{d^4f}{dH^4} \right \} + {\cal {O}}(\hbar^2) \; . 
\end{eqnarray}
Note that, as it is expected,
$f(H)=\exp (-\beta H)$ in the previous expression recovers 
the usual results\cite{6a,7a,7b}.

\section{The case of Tsallis thermostatistics}

We apply now the previous findings for Tsallis
 thermostatistics\cite{8}.
The entropy is defined as 
\begin{eqnarray}
\label{sq}
S_{q} = \frac{Tr[\hat{\rho} (1 - \hat{\rho}^{q-1})]}{q-1}\; ,
\end{eqnarray}
where $q$ is a real parameter and $\rho$ is the density matrix. 
Note that $q$ gives, essentially, the measure of the non-extensivity 
of (\ref{sq}). 
Furthermore, to thermal averages, one has to use the $q$-expectation 
value defined as
\begin{eqnarray}
\label{ave}
<\hat{A}>_{q} = Tr(\hat{\rho}^{q} A)\; ,
\end{eqnarray}
for some relevant observable $\hat A$.
Thus, the canonical distribution is given by
\begin{eqnarray}\label{26}
f(\hat H)= \left[ 1- (1-q) \beta \hat H \right]^{\frac{1}{1-q}} \; .
\end{eqnarray}
When $(1-q)\beta E_n > 1$, where $E_n$ are the eigenvalue of $\hat{H}$, 
the probability is not be positive defined quantity. 
To avoid this behavior it is usually assumed that $f(\hat{H})$.
In particular, 
the Tsallis  thermostatistics reduces to the usual one 
in the limit $q\rightarrow 1$. 

The substitution of (\ref{26}) into the previous results gives 
the partition function for Tsallis thermostatistics up to $\hbar^2$
order. 
In this case we have
\begin{eqnarray}
\label{df}
\frac{d^n f}{dH^n} = (-\beta)^n q (2q-1)(3q-2)\; ...  \; 
[(n-1)q-(n-2)] [1-(1-q)\beta H]^{\frac{nq-(n-1)}{1-q}} \; . 
\end{eqnarray}
The factor $\beta^n$ above allow us to arrive at an important 
conclusion. In fact, in  the limit $\beta \rightarrow 0$
the corrections to the classical 
limit disappear, i. e.,  the quantum corrections to the 
classical partition function are irrelevant in the high 
temperature limit. 
In this case we are supposing basically that 
the derivatives $[1-(1-q)\beta H]^{\frac{nq-(n-1)}{1-q}}$
in (\ref{z}) do not diverge. 
Note that the above conclusion is a direct consequence 
of the fact that the dependence of the function $f(\hat H)$ 
with $\beta$ and $\hat H$ appears only in the form 
$\beta \hat H$.
Thus, the previous conclusion can be applied  not only 
to the Tsallis thermostatistics but to a wide 
class of possible partition functions.

\subsection{The harmonic oscillator}

To exemplify the application of the above expressions for a 
concrete case, 
we consider  a harmonic oscillator whose Hamiltonian is 
$\hat{H} = \hat{p}^2 /(2 m) + m \omega^2 \hat{x}^2$. For this
one-dimensional case, (\ref{z}) is reduced to

\begin{eqnarray}
\label{Zoh}
Z &=& \int \frac{dp\,dx}{2\pi \hbar} f(H) \nonumber \\
&+&\hbar^2\int \frac{dp\,dx}{2\pi\hbar} 
\left\{ -\frac{\omega^2}{4}
\frac{d^2 f}{dH^2} -\frac{1}{6} \left(\frac{\omega^2}{m} p^2
+ m \omega^4 x^2 \right) \frac{d^3 f}{dH^3}
 -\frac{\omega^4}{8}x^2 p^2
\frac{d^4f}{dH^4} \right\} + {\cal O} (\hbar^2),
\end{eqnarray}
where the derivatives are obtained through (\ref{df}). 
In (\ref{Zoh})
we have to evaluate the integrals for two cases: $q<1$ and $q>1$.

Let us consider first the case of $q<1$. Since the integrands
appearing in (\ref{Zoh}) are composed of successive derivatives
obtained from (\ref{df}), the relevant integrals are of the
form
\begin{eqnarray}
\label{int1}
\int dx dy \left[ 1-(a x^2 + b y^2)\right]^{\sigma}=
 \pi a^{-1/2} b^{-1/2}\; ,
\end{eqnarray}
\begin{eqnarray}
\label{int2}
\int dx dy x^{2} \left[ 1-(a x^2 + b y^2)\right]^{\sigma}=
\pi a^{-3/2} b^{-1/2} \frac{1}{2(1+\sigma)(2+\sigma)}\; ,
\end{eqnarray}
and
\begin{eqnarray}
\label{int3}
\int dx dy x^{2}y^{2} \left[ 1-(a x^2 + b y^2)\right]^{\sigma}=
\pi a^{-3/2} b^{-3/2} \frac{1}{8(1+\sigma)(2+\sigma)(3+\sigma)}\;,
\end{eqnarray}
with $\sigma >-1$ in order to have well defined integrals. 
The above integrals are  more easily performed if
the transformation $ x= a^{-1/2} r \cos\theta$ and $y=b^{-1/2}r
\sin\theta$ is employed.
 By substituting the resulting integrals in (\ref{Zoh}) we obtain
 for the partition function
\begin{eqnarray}
\label{ZZ}
Z= \frac{1}{2-q} \frac{1}{\hbar \omega \beta}
- \frac{\hbar \omega \beta}{24} + {\cal O}(\hbar^2).
\end{eqnarray}
One observes that in the limiting case $q \rightarrow 1 $
 we recover the
usual result. Moreover, the second factor which accounts for
the quantum corrections in the first order in $\hbar$ is
 independent of
$q$ for the case of harmonic oscillator. It is important to
 stress
that the above results are meaningful for $q> 2/3$ in order
 to achieve
the convergence of all the integrals in (\ref{Zoh}). 
In the general case, 
the lower limit for $q$ is fixed by the higher order derivative
of $f(H)$ present in the expansion for $Z$.

The calculations for the case $q>1$ are similar to the 
previous one and
lead to the same expression for $Z$, i.e., (\ref{ZZ}).
 However, in this
case the upper limit is $q< 2$, and it is now
fixed by the exponent of the first term
in the expansion for $Z$.

\section{Conclusions}

 In this paper we have presented for the first time
 a general procedure to obtain all the
 quantum corrections for generalized partition functions. As
 underlined before, this procedure is a generalization of the
 Wigner expansion for usual statistical mechanics. However,
 instead to
 use Wigner  representation, we have used
 the coherent states representation, which is convenient for our
 purposes. The general formalism was used to explicitly
 evaluate the
 first quantum correction to the classical case. This result was
 applied to a non-extensive (Tsallis) thermostatistics.
 As an application
 we have considered the case of the harmonic oscillator. 
In particular,
 we have shown that the dominant term in the partition function
 depends
 on $q$. It is reduced to the usual one when $q \rightarrow 1$ as
 expected. Moreover, the first
 quantum correction for the partition function of the
 harmonic oscillator coming from the
 Tsallis thermostatistics
 is shown to be the same as the usual one. We have also determined
 the specific heat of the
 system 
 for $q>1$ and $q<1$.

\acknowledgements
One of us (R.S.M.) thanks C. Tsallis for illuminating discussions. 
This work was supported partially by CNPq (Brazil).

\end{document}